\documentclass[3p,12pt,preprint]{elsarticle}
\usepackage{hyperref}
\usepackage{setspace}
\usepackage{xspace}
\usepackage{xcolor}
\usepackage{graphicx}
\usepackage{lineno}

\title{Discrete Stochastic Models in Continuous Time for Ecology}
\author[corn]{Andrew J.\ Dolgert\corref{cor1}}
\ead{ajd27@cornell.edu}

\cortext[cor1]{Corresponding author}

\address[corn]{BRC Bioinformatics Facility, Institute of Biotechnology,
619 Rhodes Hall, Cornell University, Ithaca, NY 14853}
\begin{document}
\maketitle

\newcommand{\prob}[1]{{P}\left[{#1}\right]}

\begin{abstract}
This article shows how to specify and construct
a discrete, stochastic, continuous-time model
specifically for ecological systems. The model is
more broad than typical chemical kinetics models
in two ways.
First, using time-dependent hazard rates simplifies
the process of making models more faithful.
Second, the state of the system includes individual
traits and use of environmental resources.
The models defined here focus on taking
survival analysis of observations in the field and
using the measured hazard rates to generate
simulations which match exactly what was measured.
\end{abstract}

\begin{keyword}
agent-based modeling \sep individual-based modeling \sep process scheduling \sep non-Poisson \sep generalized semi-Markov processes
\end{keyword}

\section{Introduction}
Discrete stochastic models in continuous time are
the simplest way to turn survival analysis
of an ecological system into a simulation of that system.
They build models
most faithful to a picture of individuals making
choices which affect each other.
This article describes how rules for
individual behavior can construct a simulation of a population.
It argues that, within the context of discrete stochastic
simulation, hazard rates for transitions are
intrinsic descriptions of individual behavior.\footnote{%
Abbreviations used in this article: \textsc{mrp} for Markov renewal process,
\textsc{llcp} for long-lived competing process,
\textsc{gspn} for generalized semi-Markov Petri net, and
\textsc{gsmp} for generalized semi-Markov process.}

Discrete stochastic models in continuous time are associated
with the Gillespie algorithm and its variants, including
the Next Reaction algorithm.
The Gillespie algorithm samples to find
the next state and time of a stochastic process. The process
and sampling algorithm are separate. This article won't
define a new sampling algorithm but will examine
the stochastic process that is sampled.

Almost every discrete stochastic model in continuous
time used for ecology is based upon a chemical kinetics
model. These chemical kinetics 
models take the form of chemical species which interact
according to stoichiometry at rates specified by propensities.
They
have been enormously successful for ecological
applications such as stochastic movement
models\cite{Patterson2008,Choquet2011,Nathan2008,Durrett1994}
and any model that builds group behavior from
individual behavior\cite{Gables2014,Huston,Grimm2005}.
Clear statement of chemical kinetics as a process leads
to clear specification of chemical kinetics models
as matrices and vectors in Systems Biology Markup Language and computer code
which, in turn, leads to their utility to solve
common problems.

The chemical kinetics model constrains
expression of ecological problems in two ways.
The first is that every state in a chemical kinetics
model is a count of chemical species when an ecologist
might want the state of an individual to carry properties
that record its life history. The second is that
propensities are usually simple rates when an ecologist
might want rates to be time-dependent, so that an
insect gets more hungry over time, or a cow waits
to reproduce again, or a frog jumps quickly if it jumps early.
These two important modifications are that individuals
hold state and their hazard rates for change are time-dependent.

Two particular areas are ripe for
use of non-constant hazard rates and hazard rates
more specific to individual histories.
The first is stochastic movement models, which already have
a strong history\cite{Patterson2008,Choquet2011,Nathan2008,Durrett1994}.
The second is infectious processes.
Recent studies show that while constant hazards for infection
model well the most intense parts of outbreaks any
analysis of early outbreaks and the distribution of early
outbreaks will find incorrect $R_0$ if they do not
include Gamma-distributed
or Weibull-distributed recovery times\cite{Yan2008,Dunn2013}.

The statistical process defined in this article
includes the ability for individuals to have state
associated with them and the ability for hazard rates
to be time-dependent, and it is called long-lived
competing processes (\textsc{llcp}).
While it is known that the Gillespie algorithm can
sample such an \textsc{llcp} process, the process, itself, has not been
specified. Lacking a specification, previous uses
of time-dependent hazards and property-carrying state
have misunderstood both how models relate with observation
and how to sample the trajectory from such a model.
The bovine viral diarrhea models of Viet et al.\cite{Viet2004}
demonstrate how time-dependent hazards can include
farm management decisions within a continuous-time model,
but they also mistakenly parameterize transitions in the
model from holding times instead of hazard rates, both
of which are explained in the results section below.
The wasp model of Ewing et al.\cite{Ewing2002} calculates
competition among behavioral drives, including temperature
dependence, using a detailed continuous-time model with
time-dependent hazards, but the sampling method
used is statistically biased. In both cases, the resulting
trajectories are biased, so that summary data, such as
average survival, are incorrect. Correctness here is a
question of whether observation times and theoretical
rates put into a model are the same that come out of
simulation of the model.
Correctness is a question of whether the statistical
machinery fails the effort required to gather data and
parameterize a model.

The other consequence of lacking a specification is that
there are no ready tools with which to construct a
discrete stochastic process in continuous time which
has time-dependent hazard rates. A mathematical specification
is the first step to understanding how to state a process
clearly in a file or in code, such that it can be sampled.
It was the goal of the authors to make construction
of ecological and epidemiological models trustworthy
and rapid.

The central tenet of the stochastic processes
defined in this article is that individuals compete.
Not only do individuals compete, but also
for a single individual, the drive to eat, the drive to reproduce,
and the drive to move compete. For these models,
an individual is the sum of its drives.
Each of those drives is modeled by a separate stochastic
process whose hazard rate comes from survival analysis.
The long-lived competing process is built from a
set of conditionally-independent stochastic processes,
conditional on interruption by other processes. Each
process has its own time-dependent hazard to fire.

\begin{figure}
\centerline{\includegraphics[scale=0.4]{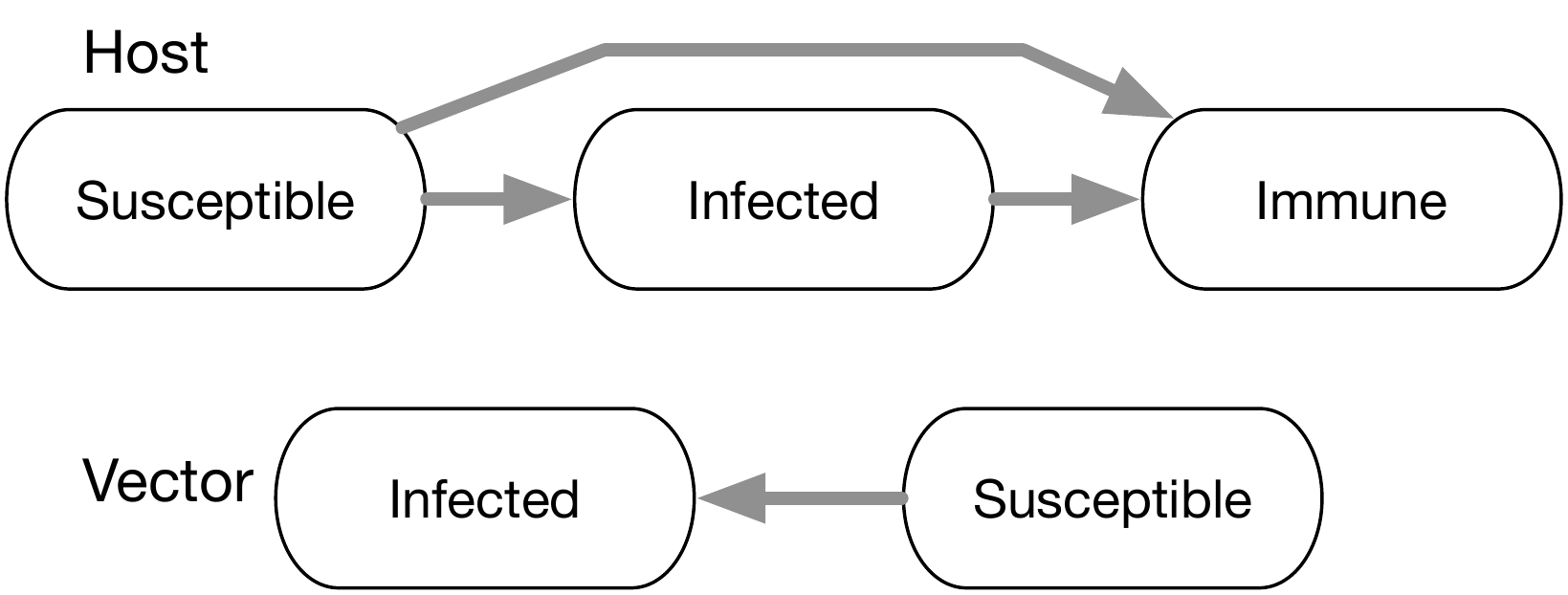}}
\caption{A simple compartmental model showing a vector
infecting a host.\label{fig:compartmental}}
\end{figure}
\begin{figure}
\centerline{\includegraphics[scale=0.4]{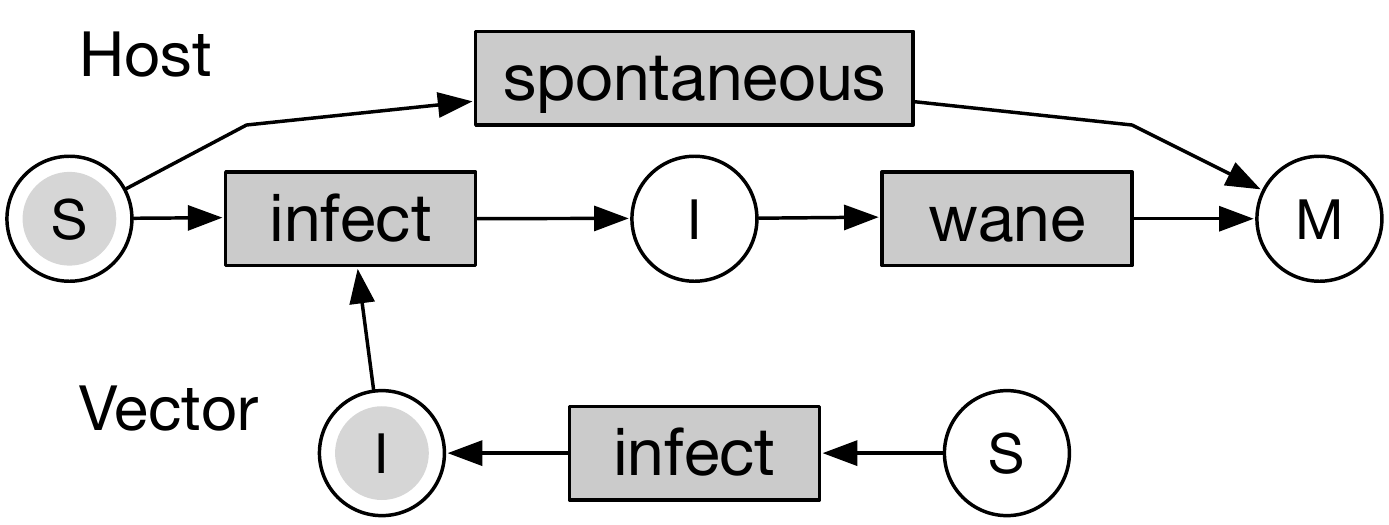}}
\caption{This diagram emphasizes transitions between
states because each transition is associated with a stochastic
process.  Host states label susceptible,
infected, and immune. Vector state label susceptible
and infected.\label{fig:compartmentpetri}}
\end{figure}%
To understand the aptness of a competing process
formulation take
a classic compartmental diagram, such as the
vector and host model in Fig.~\ref{fig:compartmental},
shows which state transitions are possible for
an individual. This diagram, given rates for transitions,
could guide construction
of a differential equation model for a population,
but think here modeling individual behaviors.
The diagram in Fig.~\ref{fig:compartmentpetri} represents
rules for how an individual can change state from one
compartment to another and at what rate the state
can change.
It emphasizes possible transitions from which move those
individuals from one compartmental state to the next.
Sec.~\ref{sec:gspn} shows how to make this diagram
into a specification for a model in the same way 
that a stoichiometric matrix is a specification for
a chemical model. 

The methods section of this article and appendix
construct a representation of a statistical process
in order to show generality and completeness.
The results section covers how to construct and
parameterize this process.

\section{Previous Work}
This article defines a class of stochastic processes
without discussing the algorithms with which to sample
trajectories of those processes. These algorithms are commonly
known as Gillespie-type and encompass the following:
Direct, Direct non-Markovian, First Reaction, Next Reaction, Anderson's
Next Reaction\cite{Gillespie2007,Gillespie1978,Gibson2000,Anderson2007}. This article doesn't extend these algorithms but
rather expands the possible ways to define changes of state, given Gillespie-type dynamics in time.

While the exposition within this article relies on
calculus of stochastic variables, a more modern approach
to discussion of Markov renewal processes uses martingales
and random time changes, as described by
Anderson and Kurtz\cite{anderson2011continuous}.
Restricting mathematical development to calculus is more
approachable but will falter to answer more precise
questions about what it means for a competing process
to be well-defined.

There is a mathematical identity that any
discrete, stochastic, continuous-time process
can be rewritten as a set of competing processes,
one for each possible next state and defined
anew at each time step\cite{Howard2007,Berman1963}. The long-lived competing processes described
below are an extension of competing processes
to encompass biological
behaviors which aren't reset each time anything
in the system changes, as happens for competing processes. It can
be considered a subset of an older technique called the
Generalized Semi-Markov Process (\textsc{gsmp}), used in
engineering\cite{Glynn1989}. The main criticisms of \textsc{gsmp}
simulations are that they 
can be complicated to specify and slow to simulate. Restriction
of the long-lived competing processes to conform to Markov
renewal processes reduces complexity somewhat and greatly speeds
computation by making possible the use of Next Reaction method.

The Generalized Stochastic Petri Net (\textsc{gspn})
in Sec.~\ref{sec:gspn} is not
only a data structure but also a set of engineering
techniques for building a process using that data
structure\cite{Conte1991,Haas2002}. Those
rules are more amenable to human-mediated processes and are
a superset of what is described here.
This article focuses on simulation as enactment of
estimations from survival analysis, so it skips over much
of the extensive feature set, and complication,
associated with use of \textsc{gspn}
in manufacturing and reliability analysis.

There have been major frameworks designed and used
for general individual-based modeling in ecology.
The \textsc{devs} discrete-event simulation is a general-purpose
simulation tool adopted for ecology\cite{Zeigler1993}.
Later groups made tools more focused on ecological
simulation in order to reduce the burden of programming
and design. \textsc{osiris} and \textsc{wesp} are good
examples\cite{Mooij1996,Lorek1999}. Both focused on
ease-of-use and correctness by combining higher-level
abstraction with software architecture.

In contrast to these frameworks,
the work presented here is a simpler mathematical
model. There is a sample implementation as a C++ library,
and references to data structures and algorithms are
sufficient to implement the model\cite{SemiMarkov2014}.
The emphasis, however,
is on how to join statistical estimation of a system
with its simulation because understanding causal
dependence of a system is more complicated than
programming, and quantitative understanding begins with
statistical correctness.

\section{Methods}

\subsection{Markov Renewal Process}\label{sec:mrp}
To call a simulation a discrete stochastic process in continuous time,
means that it is a representation of
a Markov renewal process (\textsc{mrp}). Take the \textsc{mrp} as the definition
of what it means to be such a simulation.
The structure of an \textsc{mrp}
determines in what way a model approximates the real world.
It also circumscribes limitations on what choices in a simulation
conform to mathematical requirements. The \textsc{mrp} is well-known and
this exposition follows \c{C}inlar\cite{Cinlar1975} in order to
explain how all of the actions of every individual in a simulation,
taken together, form one single \textsc{mrp} from which to sample
trajectories.

\begin{figure}
\centerline{\includegraphics[scale=0.4]{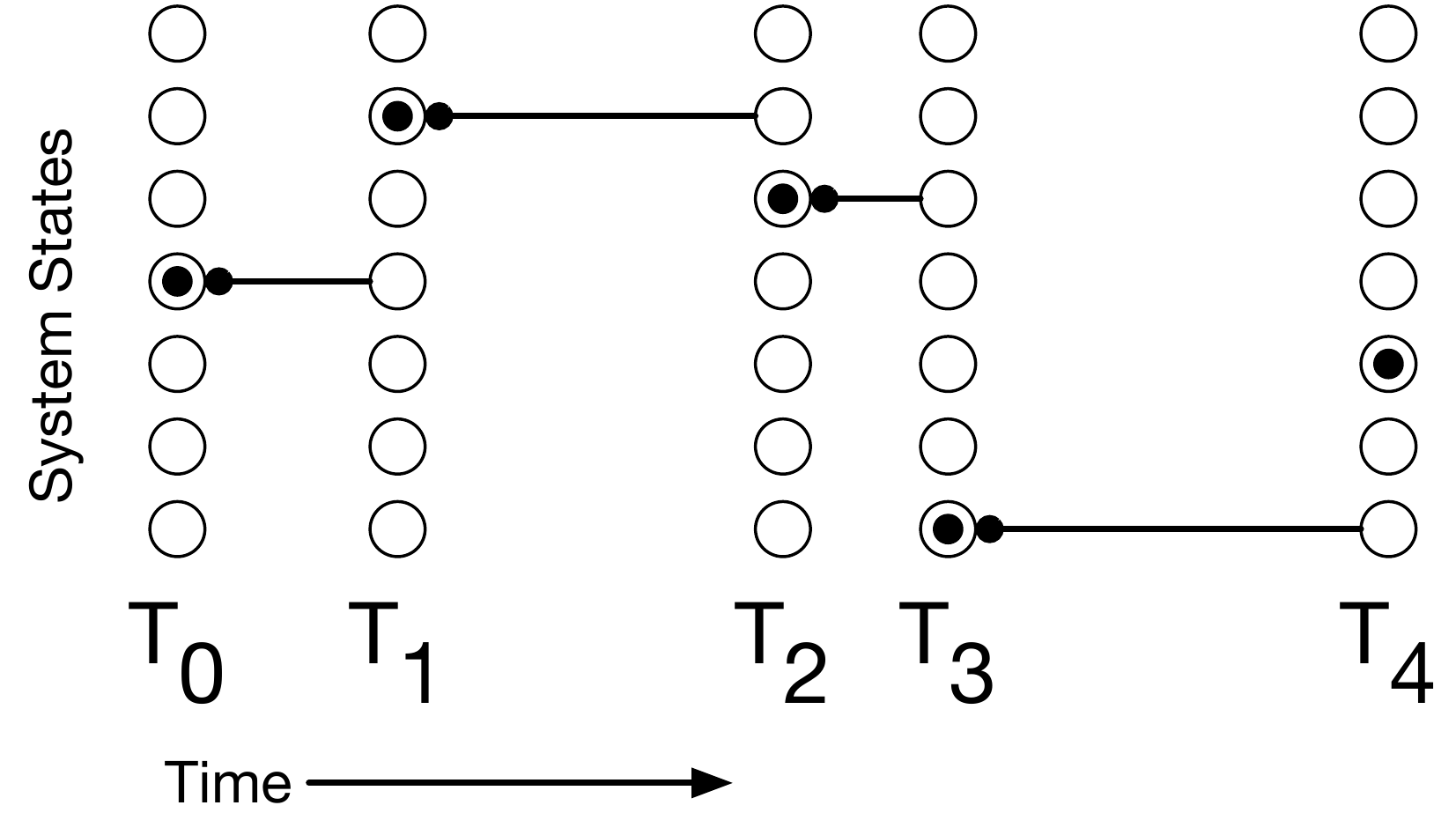}}
\caption{A Markov renewal process defines a probability
distribution of a next state at a next time given the current state.
Each circle represents a different state of the system.
Each arrow is a transition to a next state and time.
From now, what is next?\label{fig:mrp}}
\end{figure}

For an \textsc{mrp}, time is a random variable,
so it jumps from an initial time, $T_0$, to the next time, $T_1\ge T_0$.
For any jump, from $T_n$ to $T_{n+1}$, the 
state of the system is defined at those times, not in-between.
The times themselves form a countable set.
The state of the system, $X_n$, is,
itself, a random variable on a
discrete space, so it is labeled $X_n=i$.

The central engine of an \textsc{mrp} is that the next state and the next
time are chosen according to a joint probability for the next
state to be state $X_{n+1}=j$ at time $T_{n+1}$
given current state $X_n=i$ and time $T_n$,
\begin{equation}
  P\left[X_{n+1}=j, T_{n+1}-T_n\le t|X_n=i\right].\label{eqn:jointmrp}
\end{equation}
The density of this probability, its derivative, is
called the semi-Markov kernel, denoted $q_{ij}(T_{n+1}-T_n)$.
A process which conforms to an \textsc{mrp} specifies
all the next possible states of
the system, and, given the current state,
the joint probability of each next state and time
for that state.

\begin{figure}
\centerline{\includegraphics[scale=0.4]{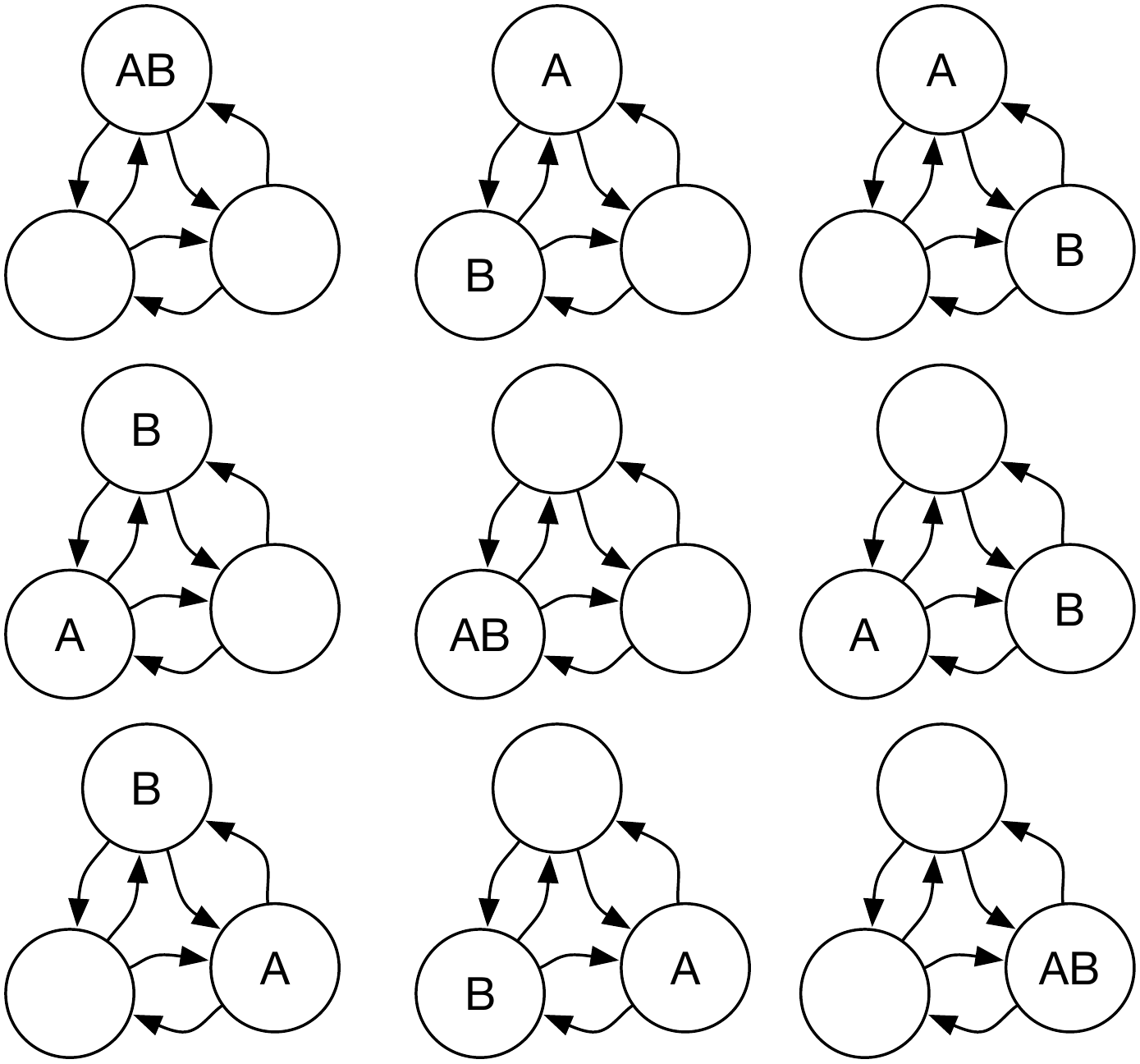}}
\caption{Given a system with two individuals, A and B,
which can move among three metapopulations, there are
nine physical states of the system, as shown. Because the
\textsc{mrp} can depend on times in a state, the complete
state of the system, $X_n$, can include the times at which
A and B arrived in their current configuration.\label{fig:configurations}}
\end{figure}

An \textsc{mrp} is a view of an ecological simulation, zoomed out
so that anything that happens, in any part of the simulation,
is the next change to the state of the whole simulation,
as shown by the states in Fig.~\ref{fig:configurations}.
Saying that a complex simulation is an instance of an \textsc{mrp}
is a statement that all of the biological and physical
processes within a simulation become a single statistical
process, marching forward in time, according to a single joint
probability distribution. Identification of that joint
probability distribution is what permits uniform
sampling in a simulation.

There is no restriction for an \textsc{mrp} that
the probability distribution for the next states be exponential.
For every Markov random process, the next state depends only
on the system state $X_n$ at $T_n$. For a Markov random process
whose distribution of next stopping times is always exponential,
the probability of arrival at a particular next state not only
doesn't depend on states before $X_n$ but also doesn't
depend on the amount of time the system takes to arrive
at that next state.

The exposition here treats the state space, $X_n$, as
a finite set of states, which would be restrictive in practice.
The number of states can be infinite as long as the next
possible states, $X_{n+1}=j$ can be sampled statistically.
\c{C}inlar's careful presentation allows states to
be real numbers, explaining that, given a discrete set
of stopping times, the set of real numbers chosen to
be stopping times is itself a finite set.

A process which conforms to an \textsc{mrp} must define a set of
initial states, a set of next states given any current state,
and a joint probability for the next states and times
at which they might occur. The next section constructs
such a process, starting with intuition for competing
processes.

\subsection{Long-lived Competing Processes}
The guiding idea for constructing this specific \textsc{mrp} is that
individuals compete for resources and modify the environment.
There will be multiple simultaneous stochastic processes,
each of which represents a tendency towards changing the
individual or the environment. Definition of
long-lived competing processes (\textsc{llcp})
proceeds in two steps, defining the state of the overall system,
and showing how to calculate the probability of the next
state and time, Eq.~\ref{eqn:jointmrp},
from current state.

\begin{figure}
\centerline{\includegraphics[scale=0.4]{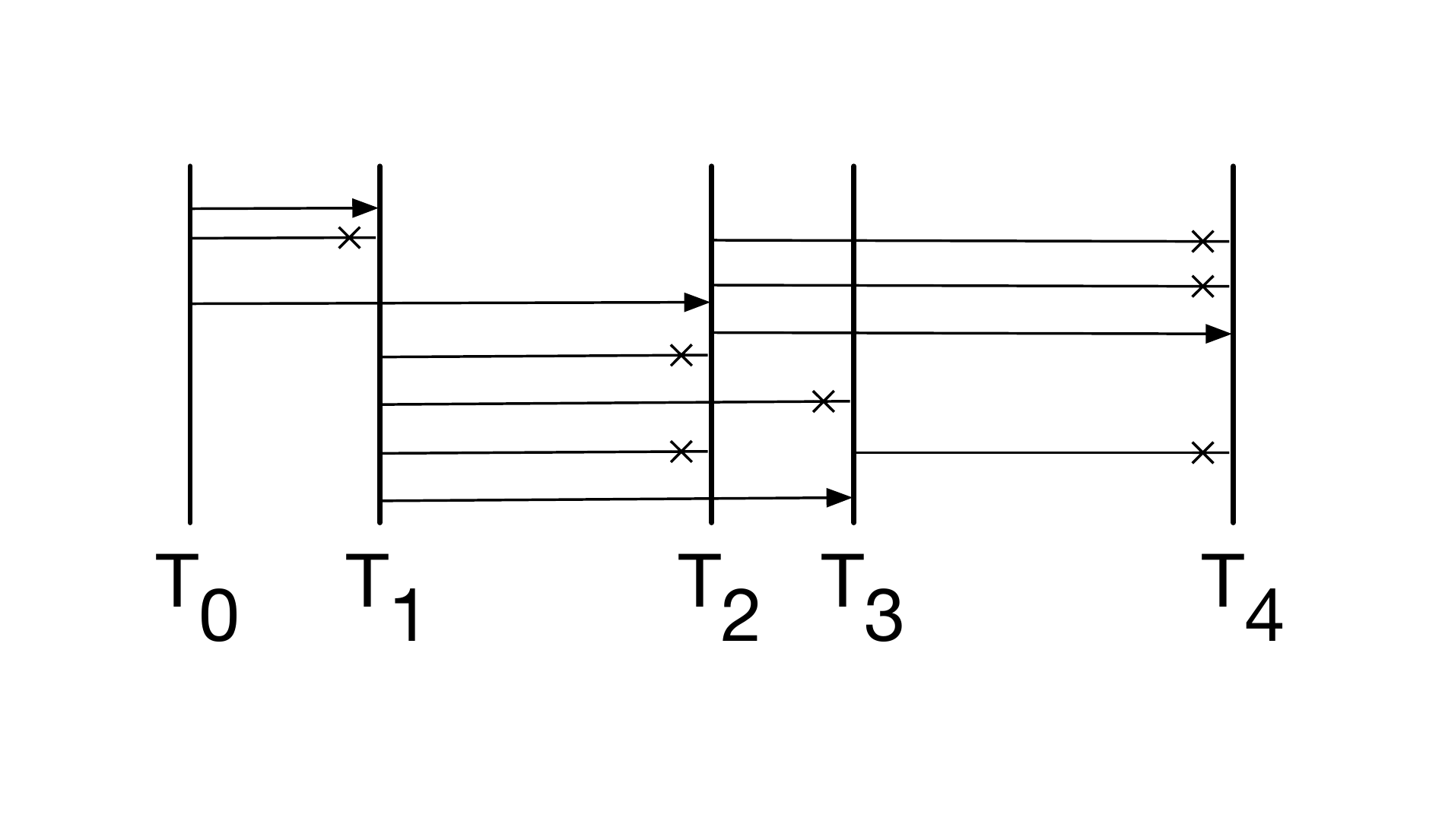}}
\caption{Long-lived competing processes define individual actions
which, together, create a Markov renewal process. Only one long-lived
process fires at a time. Others may be enabled or disabled when
one fires, but at no other time. There isn't a one-to-one
association between a competing process and the state of the system.
Each process modifies the joint substates of the system in some
known way to cause a change in the current state.\label{fig:lr}}
\end{figure}

\begin{figure}
\centerline{\includegraphics[scale=0.4]{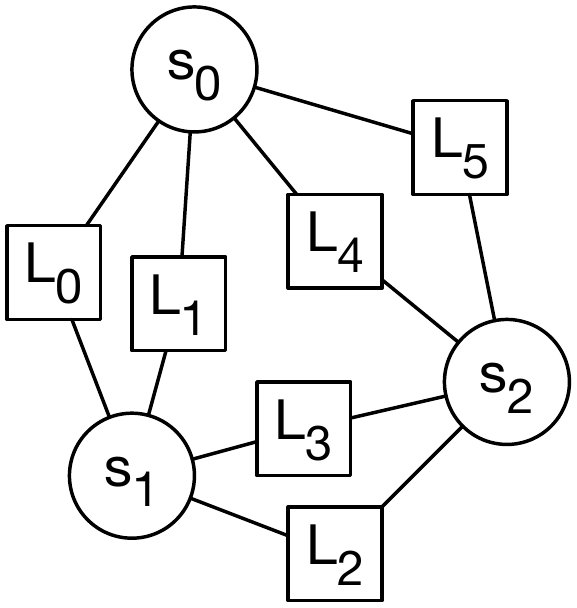}}
\caption{The same metapopulation simulation of A and B
moving among three locations could be modeled
with three physical substates, one for each metapopulation,
and six processes which compete to move an
individual to neighboring metapopulations. The competing
processes are enabled only when a substate has individuals
to move.\label{fig:lrcp-metapop}}
\end{figure}

The state of a system with many individuals, such as
the metapopulation in Fig.~\ref{fig:lrcp-metapop}, is
exponentially large (the number of trajectories given an initial value will be combinatorially large).
Any time any one
of the individuals moves, sthe whole system is in a new state, $X_{n+1}=j$.
The total number of states is the number of arrangements of individuals,
and times at which they could arrive at those arrangements.
A goal of this the \textsc{llcp} specification is to ensure
that the behavior of any particular individual can depend
on only a local environment. Whether its local environment is
its location or the its neighboring individuals,
that individual's behavior
will be independent of any behavior that doesn't interfere
with its local environment.

The \emph{physical state} of the system is the state of
individuals and their environment, separate from times
at which the system entered this physical state.
This definition follows the language of the \textsc{gsmp}
in Glynn\cite{Glynn1989}. It
is specified as a set of substates, $s_0, s_1,\ldots s_p$, so
that those substates can represent the local environment of
an individual.
This could be the location of
each individual or nearby food, for example. The substates are a disjoint
set which, together, describe the whole physical state of the \textsc{mrp}.

Each possible \emph{cause\/} for change in the system
is represented by a separate stochastic variable, $L_m$.
For instance, ``individual two jumps down'' would be a single cause
for change. At the moment the individual arrives in a spot, the
cause to jump away from that spot is
called \emph{enabled\/} at time $T_m^e$. That
cause may \emph{fire\/}, or a competing cause may \emph{disable\/}
it. For instance, the firing of ``individual two jumps right'' would
prevent that individual from having jumped North.

The probability for a cause to fire at any time depends on
some subset, $\{s_i\}$, of the substates of the physical system
and can always be written as a cumulative distribution
function defined by a time-dependent hazard rate, $\lambda_m(t)$,
\begin{equation}
  P\left[L_m\le T_m|\{s_i\}\right]=1-\exp\left[-\int_{T_m^e}^{T_m}\lambda_m(s)ds\right].\label{eq:causeprob}
\end{equation}
This is the probability for the time at which a particular
cause will happen, were it to happen. It is defined, not since the
last transition, but in absolute time, since the start of a
simulation. Associated with this cause
is a rule for when it is enabled, when it is disabled, and
what it does to the substates when it fires. Those three rules
are functions of the substates associated with this cause and
no other physical substates.

\begin{figure}
\centerline{\includegraphics[scale=0.4]{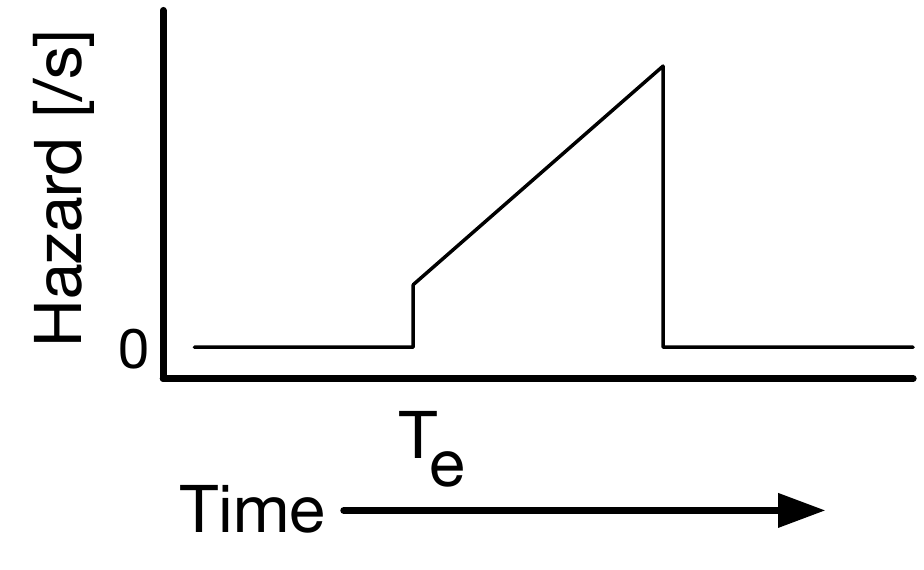}}
\caption{Long-lived competing processes are defined
in absolute time since the start of the simulation
because they represent what is allowed within a model.
When a long-lived competing process is enabled, it
has nonzero hazard. When it fires or becomes disabled,
the hazard returns to zero.\label{fig:lrhazard}}
\end{figure}
A simulation defines many causes simultaneously.
From all of these causes,
the first to fire determines what happens next in the system.
The next time, $T_{n+1}$, for the whole system as an \textsc{mrp}
is the minimum of the set of stochastic variables,
\begin{equation}
  T_{n+1}=\mbox{min}\left(\{L_m\}\right),\label{eq:mincompete}
\end{equation}
taking into account that none of the enabled causes
fired before $T_n$.
It's a competition to be the first to fire, determined
by the cause-specific probabilities for firing. When one fires,
it may disable or enable other causes.
The cause that fires determines how the system changes
to a new state, $X_{n+1}=j$.
Multiple causes may independently effect the same new state,
in which case they, together, define the
\emph{all-causes probability} of that state.
The joint probability for $X_{n+1}$ and $T_{n+1}$
is given in appendix~A.

No cause that is competing, in the sense that the firing
of another cause could disable it, will ever fire with
the probability distribution given in Eq.~\ref{eq:causeprob}.
However, the long-lived competing process guarantees that
the hazard rate of a cause, as determined by standard
survival analysis, will match the hazard rate of Eq.~\ref{eq:causeprob},
because the hazard rate is the rate at which a cause
will fire, were it to fire. To define an \textsc{llcp},

\begin{itemize}
  \item Define physical substates, $s_0, s_1,\ldots s_p$,
  \item For each cause, define
  \begin{itemize}
    \item An enabling function on the substates which
          determines enabling and, if enabled, returns
          the stochastic variable, $L_m$, and
    \item A firing function, which modifies substates.
  \end{itemize}
  \item Substates and causes form an undirected,
        bipartite graph.
\end{itemize}

Haas covers the possibility of incomplete
distributions for a similar process, the \textsc{gsmp}\cite{Haas2002}.
If there is some chance that 
the biological process associated with a cause
might never happen, even in the absence of competition,
then the probability in Eq.~\ref{eq:causeprob}
will be incomplete, meaning it will not sum to one.
Only if there is some probability that no cause in
the system fires, so that the simulation stops entirely,
will the probability of Eq.~\ref{eqn:jointmrp}
be incomplete. Most often it will sum to one or be zero
entirely when the system reaches what is called an
absorbing state. In an ecological simulation, an
absorbing state could mean all individuals are recovered
from disease, or it could be a delicate way to say
every individual has died.

A consequence of this construction of an \textsc{llcp} is that the
enabling times, $T_m^e$, for the long-lived processes become part
of the state of the system. For simple Markov processes, as opposed
to \textsc{mrp}s, those enabling times don't affect the choice of the next
state but, in general, they are important for non-exponential
distributions. While the times are real-valued, the state of
the system is discrete because those times are drawn from the
set of times, $T_n$, at which the system is defined.

Within an \textsc{llcp}, no two long-lived processes may
fire simultaneously. For instance, no two causes could both
have distributions in time which fire exactly
at $5\:\mbox{pm}$. It is possible to define a more elaborate
process which accounts for such conflict, but it would not as clearly
obey hazard rates from survival analysis and would require
a sampling mechanism more complicated than Gillespie-type.

It is possible to define an infinite number of physical states
and even an infinite number of possible transitions
(for instance, an ant wandering on sand), as long as the number
of enabled transitions is finite after each time step.
The restriction is that it must be possible to sample the
central probability, Eq.~\ref{eqn:jointmrp}.
Computationally, this would require storing only that part
of the state that is currently relevant to the simulation.

The long-lived competing process, as a pure expression
of simulation by survival analysis, can be seen as a stripped-down
version of the Generalized Semi-Markov Process (\textsc{gsmp}).
The main historical criticism of \textsc{gsmp} is that such arbitrary definition
of what is a substate and when causes are enabled or disabled
produces a slow simulation. However, solution for the next
time step, Eq.~\ref{eq:mincompete}, is exactly what Gillespie's
algorithm solves, and the tremendous speedup from
Gibson and Bruck's Next Reaction Method
relies on recording which causes depend on which physical
states. Those two details are what make an \textsc{llcp} more
computable than an arbitrary continuous-time stochastic system.

\subsection{Generalized Stochastic Petri Net\label{sec:gspn}}
A Generalized Stochastic Petri Net (\textsc{gspn}) is a family of
stochastic processes and associated data structures.
This section defines a long-lived competing process
using the data structure and associated nomenclature
of \textsc{gspn}.
Being specific about the data structure creates a clear relationship
between the firing of a process and how the state of
the system as a whole changes.


The state of a system, excluding its transitions, is defined by \emph{tokens} at \emph{places}, much like the location of checkers on a board defines the state of a game. Each place has a unique identifier.
The tokens at a place are a physical substate, $s_i$, of the
\textsc{llcp}. While chemical kinetics
simulations usually describe only a count of molecules of a species,
individuals in an ecological simulation will often have more life
history, such as birth times or subspecies, which can be represented
as properties of a token.

\begin{figure}
\centerline{\includegraphics[scale=0.5]{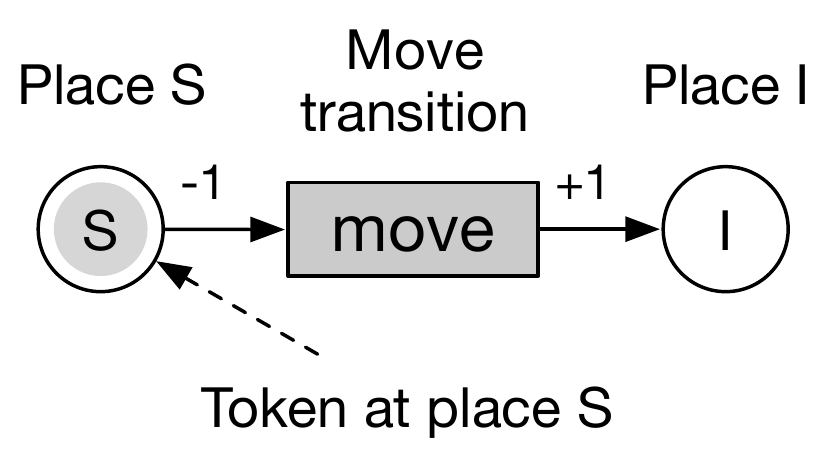}}
\caption{The language of the \textsc{gspn} is that
a transition moves tokens among places. The cardinality
of the moving tokens is stoichiometry. Each token may
have properties, which means it may contain state within the
token. This graph, accompanied by a list of properties
of the token and transition, is a specification
of a simulation.\label{fig:singletrans}}
\end{figure}

\emph{Transitions} move and modify tokens, which is called \emph{firing} a transition to produce an \emph{event}. The enabling rule
and firing rule all obey stoichiometry, which is a
count of how many tokens are taken from a place or put to a place
when a transition fires. This isn't required for the \textsc{llcp} but
has been shown by Haas to produce an equivalent statistical
process\cite{Haas2002}.
If tokens do not have properties,
then stoichiometry can be sufficient to describe both enabling and firing
rules. A transition is enabled when the number of tokens at its inputs
meets or exceeds its required inputs. It is disabled when this is no
longer true. A transition, upon firing, moves the requisite number
of tokens from inputs to outputs.

When tokens have properties, as they often will for ecological simulation,
then the specification for each transition may be augmented with
a rule such as, there must be at least one input token with the property
that the individual is older than three weeks. Hazard rates
for firing may be parameterized on properties of any tokens at 
transition inputs.

\begin{figure}
\centerline{\includegraphics[scale=0.4]{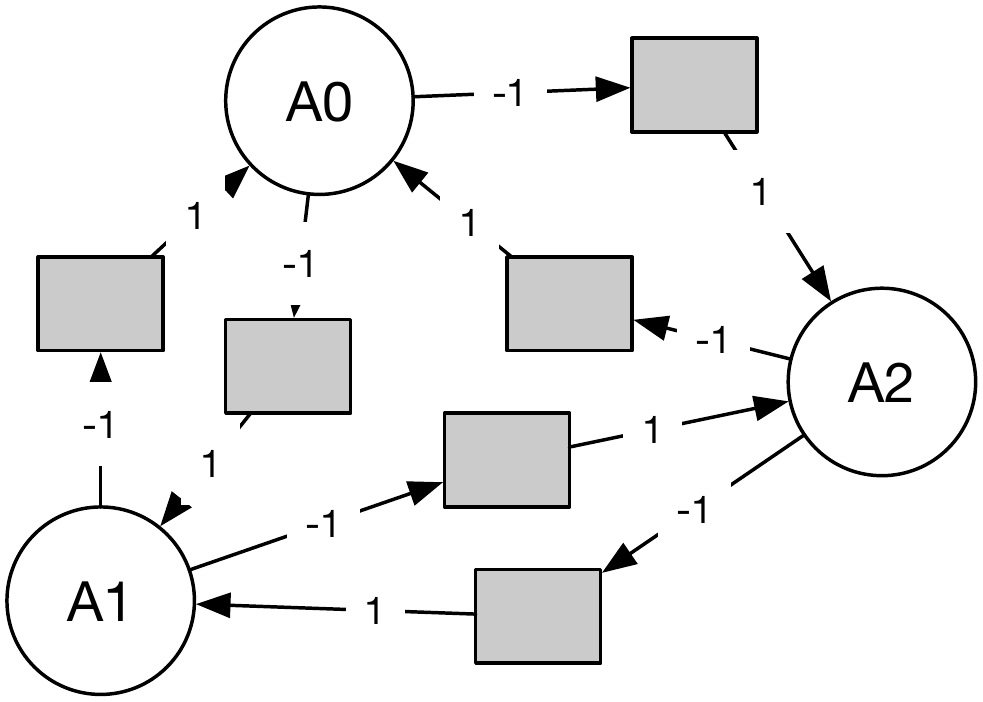}}
\caption{A movement model for a single individual moving among three
habitats might have six transitions. Coefficients on edges
indicate the number of tokens a transition needs to be enabled and
where a transition moves tokens when it fires, much like stoichiometry
in chemical systems.\label{fig:mpgspn}}
\end{figure}
\begin{figure}
\centerline{\includegraphics[scale=0.4]{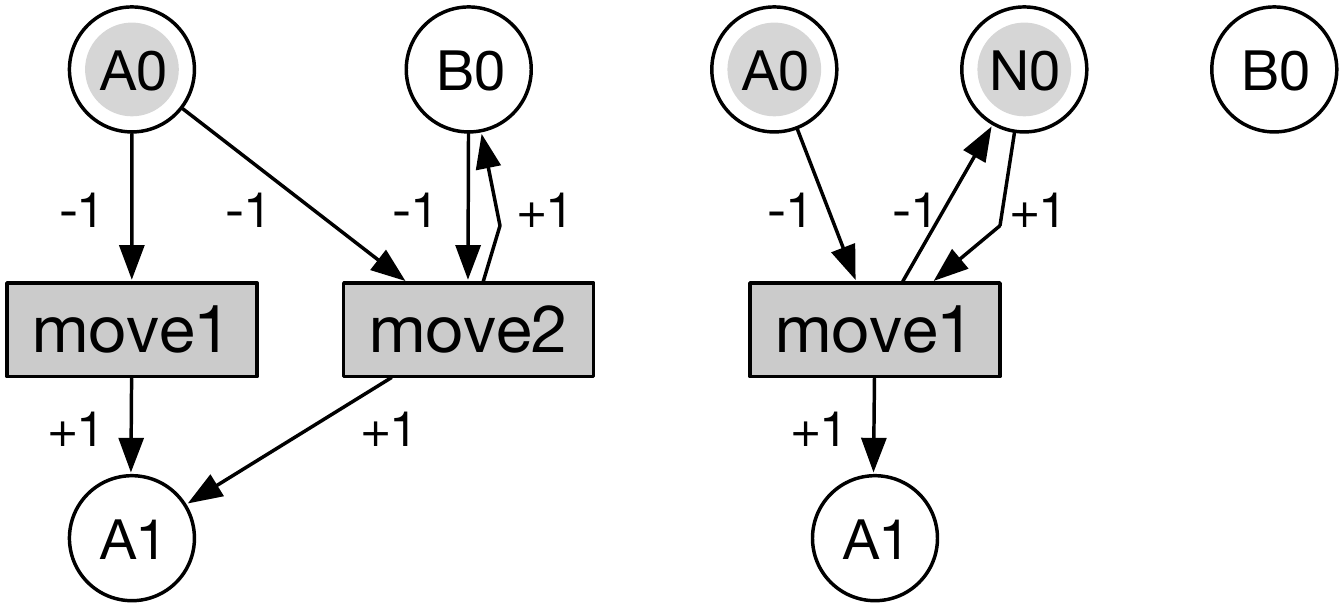}}
\caption{The stoichiometry of a \textsc{gspn} requires
movement of tokens. Shown here are two different ways to
implement a metapopulation movement model which incorporates
movement dependence on population size. The \textsc{gspn}
on the left defines a separate transition for each configuration
of occupancies. The transition ``move2'' defines a hazard
for moving individual A when individual B is present.
The \textsc{gspn} on the right creates
a separate place whose token count represents the number
of individuals in the metapopulation. This extra state
avoids combinatoric explosion of the number of
transitions.\label{fig:metapopmove}}
\end{figure}

The state of the system is not only the tokens, with their properties and locations, but also the set of \emph{enabling times} for transitions. Because this model permits non-constant hazard rates, the enabling times are necessary to know the distribution of future firing times for the system of transitions.

To define a \textsc{gspn},
\begin{itemize}
  \item Define a set of places.
  \item Define tokens which can move among those places
        and may have properties.
  \item Each transition has
  \begin{itemize}
    \item An enabling function which returns whether
          the transition is enabled on its places
          and returns the stochastic variable
          for the firing time of this transition.
    \item A stoichiometric coefficient specifying how
          many tokens are taken from or sent to each
          place upon firing.
    \item A firing function which moves and modifies
          properties of tokens.
  \end{itemize}
  \item Places and transitions form a bipartite graph
        which is directed by stoichiometry.
\end{itemize}

The \textsc{gspn} is a network in that transitions connect to places
which connect to other transitions. For most simulations,
this network can be created as a graph structure with
marking at the places and transition enabling times at
the transitions. The data on this graph is then the state
of the model. How a simulation chooses what to fire next
is separate from the state of the model, and any one of the
Gillespie algorithms would yield the same ensemble of
results.

\section{Results}

This section discusses rational construction of
such a stochastic process from real-world observations
or from mental models of ecological processes.
Consistency with the outside world can be subtle
anywhere in a model that two individuals might make
conflicting decisions, which is the point
of greatest interest.

\subsection{Choice of States}
An early step in building a model is to decide what
can be the states of the system. The granularity
of a model is a measure of how many behaviors are grouped
together. For instance,
an individual-based model of movement
might include each wandering ungulate, whereas a
more coarse-grained model could represent a single
state for the location of a whole herd.

The term ``compartmental model'' comes from the
task of distinguishing types of tissue in microscope
images. The task is assignment of a continuous observable
to a discrete class. A state in a \textsc{gspn} can be tied
to a differentiating observation or to a theoretically
relevant tipping point, such as survival of a metapopulation
past early stochastic die-off.
Something as simple as a state that represents
a location still involves choice.
For instance, a movement model
of butterflies going bush to bush might identify a state
according to the nearest bush, but it might identify a state
as the last bush on which the butterfly landed, even if
the butterfly has since taken off from that bush.
The latter choice would correspond to field observations
which record times at which insects land.

The second step is to choose how tokens on places represent
those states.
The combination of token and place together define the state.
A movement model could have a place for each location, or
it could store the current place as a property
of an individual's token. Both choices are valid, but
a \textsc{gspn} which has fewer transitions
connecting to the same place tends to be more efficient
for simulation. Two transitions which depend on the same
place equate to two biological processes which depend
on the same local environment. They are coupled, and
that coupling means a simulation has to do more work
to ensure the currency of each process when shared
state changes.

However states are chosen,
the choice of which states to define combines with
the ecological reality of which transitions among states
are possible to create a necessary set of transitions
to parameterize.

\subsection{Holding Times and Waiting Times}

The behaviors of an individual become a set of
related transitions in a \textsc{gspn}. 
Consider an example
of an individual which arrives at a metapopulation
and can move to one of four other metapopulations.
There are four transitions, one for each movement,
and all have the same enabling time, the moment
the individual arrived at its starting location.
A token starts in one place, and each transition,
were it to fire, would move it to a different place.
Together, four transitions define the movement behavior of a
single individual.

There are two questions to answer about such a system:
what happens next and when it happens. These together
are the joint probability of Eq.~\ref{eqn:jointmrp} which
is $P[\mbox{what}, \mbox{when}]$.
As with any joint probability, there are two factorizations.
These factorizations are called the holding time
and the waiting time.

The holding time formulation asks, given what happens next,
when will it happen? This factorizes the joint probability as
$P[\mbox{what}]P[\mbox{when}|\mbox{what}]$.
Two quantities together specify the density of this distribution,
the time-independent stochastic probability density, $\pi_{ij}$, of one outcome or the other,
and the time-dependence of when that outcome happens, $h_{ij}(T-T_e)$,
given that it will happen. This latter quantity is called the holding
time.
Time for a distribution is measured as time since enabling,
which is time since the token arrived at the current place.

The waiting time formulation asks, given when the next event
happens, what will happen? This factorizes the joint probability
as $P[\mbox{when}]P[\mbox{what}|\mbox{when}]$.
The density of times for the next event are called
the waiting time, $w_{i}(T-T_e)$.
This information, were it observed by itself in an experiment,
would be insufficient to determine hazard rates for transitions.
It must be augmented by the 
time-dependent stochastic probability, $\pi_{ij}(T-T_e)$,
which asks the probability of one outcome or another
at the given firing time.

Neither the holding time nor the waiting time are what go into
the \textsc{gspn}. Given measurement or specification of holding times
or waiting times, survival analysis reduces them to hazard
rates, $\lambda_{ij}(T-T_e)$, which are the probability per
unit time that an event happens, given that it has not yet happened.
It is this hazard rate that specifies the distribution of a transition in
Eq.~\ref{eq:causeprob}. For the simple example of
a token moving to one of some number of places, the hazard rate
can be found from the semi-Markov kernel written either as
$q_{ij}(T-T_e)=\pi_{ij}h_{ij}(T-T_e)$ or as
$q_{ij}(T-T_e)=w_{ij}(T-T_e)\pi_{ij}(T-T_e)$. From that
kernel, the hazard rate is
\begin{equation}
  \lambda_{ij}(T-T_e)=\frac{q_{ij}(T-T_e)}
  {1-\sum_j\int_{T_e}^Tq_{ij}(s)ds}.
\end{equation}
This simplified analysis of an individual's choices in
the presence of an environment provides states and hazards
with which to construct a system of many individuals sharing
an environment.
For observed data,
a statistical estimator, such as Nelson-Aalen or Kaplan-Meier,
construct hazards and integrated hazards which
properly account for the kind of censoring that looks,
in a \textsc{gspn}, like one transition disabling another transition\cite{Datta2001}.

The hazard rate for every transition can be stated
equivalently by a transition distribution,
$f(t)=\lambda(t)\exp(-\int \lambda(s)ds)$. The holding time, waiting time,
and transition distribution are three different
probability densities in time.
Mistaking holding time distributions for
transition distributions will construct a \textsc{gspn} whose timing
isn't what was measured or estimated.
If and only if a transition can never be disabled by another transition, so that it is independent of all other transitions, will this probability distribution be the same as the holding time which will, in turn, be the same as the waiting time. In all interesting cases, they differ.

Construction of a stochastic, continuous-time model
can be a balancing act because observed rates
for transitions, the holding or waiting times,
are the result of competition among
individuals and other ecological processes, whereas
simulation is controlled by hazard rates.
Consequently, a simulation constructed from seemingly
reasonable choices of hazard rate may show transition
rates above or below expectations because of dependency
among transitions as they compete. If some transition
rates are estimated or inferred, they may be reparameterized.
The common technique of establishing a parameter which is
a scaling factor on the transition distribution may
not be desirable because it scales hazard rate in a
time-dependent way and because it changes how incomplete
the transition is. Incompleteness implies a transition
would never fire, even in the absence of competition,
which may not be an intended consequence of rescaling.

\subsection{Markovian Systems}
The chemical kinetic models that are most commonly used
for ecology and epidemiology are Markovian, which 
implies that every transition's distribution is exponential.
Exponential distributions have the unique property that the
hazard rate is constant from the enabling time onward, so that
once a transition is enabled, it is no longer important to remember
when it was enabled. A \textsc{gspn} model whose transitions are all
exponentially-distributed therefore need not include enabling
times in its state.

Consider two individuals in the same location of
a metapopulation model. One \textsc{gspn} representation represents
their presence in a location by two tokens on two places.
Transitions can move those tokens to the other locations.
If both have identical transition rates for moving to other
locations, then an alternative representation in a \textsc{gspn} is
to place both tokens at the same place. The transition's
hazard rate is then the sum of the hazard rates of each token,
but, at the moment the transition fires, it picks only one
token to move. When all hazard rates are constant, this technique
of coalescing multiple places and transitions into one place
and one transition
can greatly speed computation. It also works, however, for
time-dependent hazard rates, as long as all coalesced hazard rates
have the same time dependence, with the same enabling time.

While an \textsc{llcp} on a \textsc{gspn} and a Markovian chemical kinetic model
both use exponential distributions, the \textsc{llcp} permits
more general rules for when transitions are enabled and
how state changes when it fires.

\subsection{Model as a Graph}
No particular representation of a model as places, tokens, and transitions is unique. There can always be another model with different places, tokens or transitions whose trajectory is one-to-one equivalent. For instance, we might represent a single insect as a token hopping among places, or we might represent it as a token with a property which is its current location. Nevertheless, the form of these models, as graphs whose nodes can be enumerated and cataloged, offers a unique chance to construct models incrementally and compare results.

Given a particular \textsc{gspn} representation of a model, every aspect of the model
is cataloged as a place, transition, or marking. When comparing
another, similar model, some of the places, transitions and markings
will be the same, and some will be different.
It is possible to make a list
which can be verified computationally. This is a significant
improvement over more free-form code.

Composition of models becomes composition of graphs. This might
be a hierarchical composition, as when individuals with disease
states are composed by a contact graph, or it could be a composition
of an ecological model with an economics model. In either case,
modular composition of Petri Nets by transition fusion or place
fusion is already described\cite{Christensen2000,Gomes2005}.
Graph fusion techniques offer a formal method for joining
\textsc{gspn}.

The \textsc{gspn} graph itself encodes causal dependency in a simulation.
Connections between transitions and places form a bipartite graph. Enforcement of stoichiometry means that every transition takes tokens from input places and gives tokens to output places, so that edges of this graph are directed.
Any transition whose firing would disable another transition makes
that other transition dependent. Stoichiometry records when removing
a token disables a transition, so it encodes this statistical
dependency. The converse statement is that transitions whose parent places are not shared cannot be dependent.

The most immediate advantage of a clear dependence structure is computational efficiency. Because the directed bipartite graph limits possible causes and effects of any firing transition, the same graph enables efficient implementation of Gibson and Bruck's Next Reaction algorithm or Anderson's method, which is an equivalent algorithm based on hazard rates. Both of these optimizations of Gillespie's
First Reaction method augment the stochastic process described
here with added state describing a remaining integrated hazard
for sampling each transition. Maintenance of this state imposes
a requirement carefully ensured for long-lived competing processes,
that any transition whose hazard rate or enabling changes when
another transition fires must be dependent on that transition.
For the \textsc{gspn}, this means it must stoichiometrically share an input
place with the transition's output places. An alternative implementation
could separate this dependency list into its own directed graph
among transitions.

\section{Conclusions}
Chemical kinetics simulations with exponentially-distributed
transitions are well understood and commonly in use
for ecology. The addition
of time-dependent hazards and more general firing rules,
as described here,
introduces significantly more complexity for theory and implementation,
but it allows design of simulations which more accurately obey
both measurement and intuition. Individual behavior
is neither measured nor imagined as happening at constant rates,
so the ability to model with time-dependent hazard rates
avoids the more complicated question of what behavior
a simulation represents.

By offering a definition of an underlying stochastic process,
this article sets ground rules so that it becomes possible
to focus on ecological questions about a simulation.
One complaint about individual-based
models is that the only specification for complex
models is code\cite{Grimm2006,Grimm2010}, and
specification of an \textsc{llcp} is entirely separate
from the code used to sample the process.
Since hazard rates are central to simulation of an \textsc{llcp},
survival analysis for complex systems becomes crucial.
Techniques borrowed from survival analysis, such as
proportional hazards and frailty models, provide
known tools for estimation of parameters for a simulation.

As noted in Grimm et al, a simulation requires a hierarchy
of choices\cite{Grimm2005}. The stochastic process supports
choice of who or what is included in the simulation and,
above that, the behaviors under study. These, in turn,
are mirrored in layers of software.
The authors provide a sample implementation of \textsc{llcp}
as a C++ library\cite{SemiMarkov2014}. This, in turn,
supports construction of an application which expresses
an ecological model.
The \textsc{llcp} is a statistical process upon which to build
modeling tools more focused on specific tasks, such
as animal movement, epidemiology, and metapopulation analysis.

The authors thank David J.~Schneider and Christopher R.~Myers
for helpful discussions.
This work was supported by the
Science \& Technology Directorate, Department of Homeland
Security, via interagency Agreement No.~HSHQDC-10-X-
00138.

\appendix
\section{Derivation of Long-lived Competing Processes}

In order to affix notation, we define the continuous
probability distribution of a non-negative, real-valued
random variable $X$, denoted $F_X(T)$, as
$F_X(T)=\mathcal{P}[X\le T]$. Given that $F_X(T$ is sufficiently
smooth, its density is the derivative, $f_X(T)=dF_X(T)/dT$.
The survival is the complement of the probability distribution,
$G_X(T)=1-F_X(T)$. Lastly, the hazard rate is
the ratio of density to survival, $\lambda_X(T)=f_X(T)/G_X(T)$.
The hazard rate specifies the density
\begin{equation}
  f_X(T)=\lambda_X(T)\exp\left[-\int_0^T\lambda_X(s)ds\right]
\end{equation}
for differentiable probability distributions.
Subject to the caveat of differentiability,
there is no loss of generality in describing stochastic processes in
terms of hazards.

A long-lived competing process defines a physical state $s_i$,
composed of substates, $s_i=(p_0, p_1,\ldots)$. There are
stochastic variables, $L_k$, defined in absolute time, $T$, since
the start of the simulation. The system state, $J_n$,
consists of both the physical state, $s_i$, and the enabling
times of each of the $L_k$. The stochastic variables,
their enabling rules, and their firing rules, form an operator
on the state of the system. This operator, distinct from
an initial state, is the model for all of what can happen
in a simulation.

At the moment of enabling a competing process, the
survival distribution has the expected form,
$G_{L_k}(T)=\mathcal{P}[L_k> T]$. Once any process has fired,
those $L_k$ which remain enabled are now known not to have
fired, so their probability distributions are rescaled.
Defining a long-lived competing process by its hazard rate,
the rescaled survival distribution integrates over future times,
\begin{equation}
    G_{L_k}(T,T_0)=\exp\left[-\int_{T_0}^T \lambda_{L_k}(s)ds\right],
\end{equation}
where $T_0$ is the current time of the last observed event.
Each of these $L_k$ has a density defined
in terms of its hazard in absolute time,
$f_{L_k}(T)=\lambda_{L_k}(T)\exp\left[
  -\int_{T_0}^T\lambda_{L_k}(s)ds\right]$
The joint density for these independent processes is their product,
$f(T)=\Pi_k f_{L_k}(T_k)$.

The probability at time $T_{n-1}$ of any next state, $J_n$ and time $T_n$ is a question
of which of the $L_k$ fires first and when it fires.
The time at which a particular process would fire, were it
first to fire, corresponds to integrating the joint density over
all ways in which every other process can fire later,
\begin{equation}
  q_{ij}(T)=\int_{T_n}^\infty\cdots\int_{T}^\infty \Pi_k f_{L_k}(s_k)ds_k,
\end{equation}
where the integrals are over all but one of the stochastic processes.
This quantity is called the semi-Markov density and, in terms of
hazard rates, integrates to
\begin{equation}
  q_{ij}(T)=\lambda_{L_k}(T)\exp\left[-\int_{T_{n-1}}^{T}\sum_k\lambda_{L_k}(s)ds\right].
\end{equation}

The semi-Markov density, $q_{ij}(T)$, is a joint discrete and continuous
probability density. It will be incomplete if any of the stochastic
variables $L_k$ are incomplete. This joint probability distribution
can be split two ways into a marginal and conditional.
Integrating the semi-Markov kernel over all time yields a stochastic
matrix, $\pi_{ij}=\int_{T_{n-1}}^T q_{ij}(s)ds$, which is the
marginal probability of choosing
a cause associated with a particular $L_k$. If all causes which
lead to the same final state are grouped together, then $p_{ij}$
is the usual stochastic matrix used to determine the
next state of the system. The conditional of $p_{ij}$ is 
called the holding time, $h_{ij}(T)=q_{ij}(T)/p_{ij}$.
This is the probability distribution for firing of a cause
given that this cause will be the next to fire.
Confusion of $h_{ij}(T)$ and $f_{L_k}(T)$ will lead to trajectories
which do not match input values.
The other marginal distribution is the waiting time until the next
event, $w_{ij}(T)=\sum_j q_{ij}(T)$, obtained by summing over all
causes which are enabled.
The marginal for this is a time-dependent stochastic matrix,
$p_{ij}(T)=q_{ij}(T)/w_{ij}(T)$.

\bibliographystyle{elsarticle-num-names}
\bibliography{specification}
\end{document}